\documentclass[11pt,a4paper]{article}%
\usepackage{amsmath}%
\usepackage{amsfonts}%
\usepackage{amssymb}%
\usepackage{graphicx}
\usepackage[margin=25mm]{geometry}

\begin{document}

\title{Technical Report\\\textbf{An Attempt of Adaptive Heightfield Rendering\\with Complex Interpolants Using Ray Casting}}
\author{Daniel Cornel\thanks{cornel@vrvis.at}, Zsolt Horváth, Jürgen Waser
\\VRVis Forschungs-GmbH, Vienna, Austria}
\date{January 26, 2022}
\maketitle

\begin{abstract}
In this technical report, we document our attempt to visualize adaptive heightfields with smooth interpolation using ray casting in real time.
The performance of ray casting depends strongly on the used interpolant and its efficient evaluation.
Unfortunately, analytical solutions for ray-surface intersections are only given in the literature for very few simple, piece-wise polynomial surfaces.
In our use case, we approximate the heightfield with radial basis functions defined on an adaptive grid, for which we propose a two-step solution:
First, we reconstruct and discretize the currently visible portion of the surface with smooth approximation into a set of off-screen buffers.
In a second step, we interpret these off-screen buffers as regular heightfields that can be rendered efficiently with ray casting using a simple bilinear interpolant.
While our approach works, our quantitative evaluation shows that the performance depends strongly on the complexity and size of the heightfield.
Real-time performance cannot be achieved for arbitrary heightfields, which is why we report our findings as a failed attempt to use ray casting for practical geospatial visualization in real time.
\end{abstract}

\section{Introduction}

Heightfield rendering is an integral component of geospatial visualization in 3D to enable the interactive display of grid-based data such as digital elevation models (DEMs) and environmental simulation data.
As the discretized height values defined on the grid do not have a visual representation of their own, a continuous surface has to be reconstructed from them by interpolation or approximation.
Various approaches have been proposed to efficiently render heightfields using linear interpolation.
Triangulation-based approaches aim at generating view-dependent meshes from heightfields, mostly tailored to terrain rendering.
Popular approaches include geometry clipmaps~\cite{LosassoEtAl2004}, projected grids~\cite{Johanson2004}, and using hardware tessellation~\cite{Bonaventura2011}.
These approaches are indirect in that they require to create a finite triangulation that approximates the continuous surface of the interpolant for efficient GPU processing.

Direct visualization of heightfields is possible with ray casting and ray marching methods.
These methods rely on the ability to find the intersection of a ray with the surface of an interpolant quickly and accurately.
Tevs~et~al. use maximum mipmaps, a quadtree structure for the optimized search of ray-surface intersections~\cite{TevsEtAl2008}.
Based on this method, Dick et al. propose an improved ray traversal and a simplified determination of the ray-surface intersection~\cite{DickEtAl2009}.
In the work of Feldmann and Hinrichs, level-of-detail-dependent early ray termination is achieved by means of clipmaps~\cite{FeldmannAndHinrichs2012}.
The method of Lee~et~al. accelerates ray casting by skipping empty spaces in the quadtree prior to the actual ray marching phase~\cite{LeeEtAl2016}.

All of the methods mentioned above use at most second-order interpolation for surface reconstruction.
However, previous work~\cite{Kidner2003} has shown that for a proper reconstruction of DEMs, higher-order interpolation is required, which is more costly to implement and evaluate.
For storage efficiency, it is also common to define heightfields on an adaptive grid such as a quadtree, which further complicates surface reconstruction.
Apart from higher-order polynomial interpolators such as third-order interpolation on a quadtree~\cite{CornelEtAl2019}, suitable approaches include approximations such as kriging~\cite{Rees2000, Cheng2013, GutierrezEtAl2014}, local refinable splines~\cite{DengEtAl2008, BrovkaEtAl2014, LiEtAl2016}, natural neighbor interpolation~\cite{Bobach2008, BeutelEtAl2010}, and radial basis functions such as thin-plate splines~\cite{Franke1982, Hutchinson1995, BeatsonEtAl2014}.

For none of these adaptive higher-order interpolation and approximation methods, an analytical ray-surface intersection calculation has been formulated, and it is known to be impossible in some cases.
In previous work, this has been alleviated by resorting to second-order interpolation only, or by employing costly iterative ray traversal algorithms to approximate the ray-surface intersection.

Instead, we propose a two-step solution that decouples higher-order surface reconstruction from rendering.
We first evaluate the surface with the help of Gaussian radial basis functions that we then discretize into a set of regular textures for caching.
To control the discretization error and guarantee high resolution in the region near to the camera, we use cascaded textures inspired by parallel-split shadow maps.
The discretized smooth heightfield in each of the textures is then rendered in a second step using ray casting with empty space skipping~\cite{TevsEtAl2008} with common ray-bilinear patch intersections.
The proposed two-step solution allows for the use of arbitrary surface reconstruction with ray casting for rendering at the cost of an additional view-dependent surface discretization pass.

\section{Overview}

\begin{figure*}[t]
	\centering
	\includegraphics[width=\linewidth]{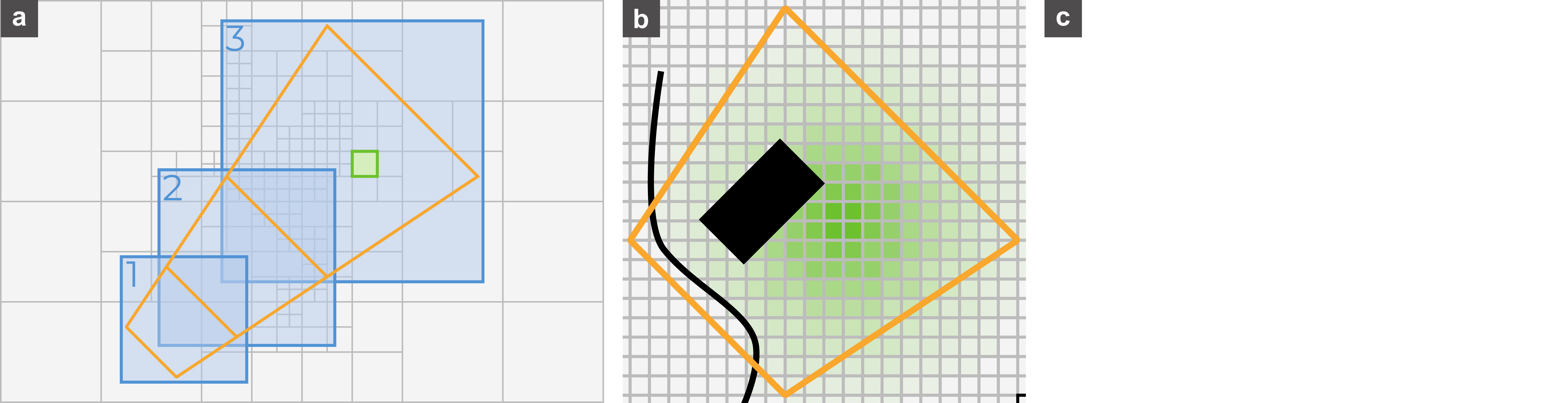}
	\caption{Overview of our technique. (a) Subdivision of the view frustum~(orange) into cascades. Each cascade is fitted into a texture~(blue). A single cell of the underlying adaptive grid is highlighted~(green). (b) Close-up view of the third cascade fitted into a texture. Smooth heightfield values are approximated by the weighted sum of Gaussian radial basis functions. The discretized Gaussian~(green) of the green cell highlighted in~(a) is illustrated. (c) Rendering of the smooth heightfield surface.}
	\label{fig:overview}
\end{figure*}

Our technique is able to use even coarse heightfield data to generate visually appealing renderings without misleading artifacts.
In our use case of a decision support system used in the context of flood management, the input data consist of a digital elevation model and relative water depths defined at the cell centers of an adaptive grid.
We define a smooth approximation of these values through the weighted sum of Gaussian radial basis functions, where each cell is given a certain radius of influence depending on its size.
We cache the approximated values in textures to enable fast repeated access to them during ray casting.

Our approach to levels of detail is inspired by parallel-split shadow maps~\cite{ZhangEtAl2006}.
By splitting the potentially visible area of the spatial domain into cascades~\cite{Engel2006} with each subsequent cascade covering a larger region than the previous one, we adjust the sampling frequencies of positions in the spatial domain according to their camera distance.
This process is fully covered in~Section~\ref{sec:cascade_fitting}.
The orange area in~Figure~\ref{fig:overview}a represents the potentially visible area of the heightfield split into three cascades.
The blue squares represent the textures in which the approximated values for each cascade will be cached.
Figure~\ref{fig:overview}b illustrates the discretization of the third (largest) cascade into a texture.
The approximation of values with radial basis functions described in Section~\ref{sec:sampling} is evaluated for each texel of these textures.
The Gaussian radial basis function of the green adaptive grid cell highlighted in Figure~\ref{fig:overview}a, discretized at each texel, is illustrated in Figure~\ref{fig:overview}b.
The rendering in Figure~\ref{fig:overview}c shows the result of ray casting the final weighted sum of the data with bilinear interpolation.

\section{Cascade Fitting}
\label{sec:cascade_fitting}

The computational costs of complex adaptive interpolation and approximation methods such as radial basis functions hinder the on-the-fly computation of values upon every request in a shader.
As the same values are needed multiple times in different stages of the visualization pipeline, a caching strategy is required.
Both time and memory limitations make it impossible to compute the smooth approximation of the heightfield for the entire spatial domain sampled at sufficiently high resolution and store it in memory.
Consequently, we only approximate the data in the potentially visible area of the spatial domain, which usually results in a much smaller data set.
However, the extents of this area can still be too large to be stored in memory at high resolution.
Therefore, we subdivide the area into several view-dependent parts that we sample at different resolutions.

Our approach is inspired by parallel-split shadow maps~\cite{ZhangEtAl2006}, which reduce perspective aliasing in shadow generation by introducing cascades.
They provide shadow maps with different sampling densities for areas at different distances from the camera.
For rendering without visible artifacts, areas closer to the camera require a higher sampling density than those in the distance.
We subdivide the volume given by the view frustum into three parts to cache the approximated heightfield data of the potentially visible area in three cascade textures.
This subdivision is illustrated in Figure~\ref{fig:overview}a, where the blue squares $1$, $2$, and $3$ indicate the three cascade textures.
All these textures have the same amount of texels.
The nearest cascade $1$ only covers a small area and provides the best resolution, whereas the farthest cascade $3$ covers the largest area at the coarsest resolution.

In order to provide data for the entire visible area, we need to align and scale the cascades within the textures in a way that each position within the visible area is covered by at least one texel in any of the three cascade textures.
Areas close to the split between each two cascades must even be covered in both corresponding textures.
This allows us to blend values at cascade transitions to avoid discontinuities.
To align the cascades, we first calculate the intersection volume of the view frustum with the bounding box of the height field.
The vertices of this volume are orthographically projected onto the ground plane and a convex hull polygon is created from the resulting points.
This hull represents the potentially visible area of the spatial domain, which is the area we want to cover with the cascades.

For the hull polygon, we determine the vertices with the lowest offset~$n$ and highest offset~$f$ along the normalized two-dimensional projection of the view vector.
These two offsets define the depth range the cascades need to cover.
We divide this range twice with logarithmic splits according to Zhang et al.~\cite{ZhangEtAl2006}, i.e., at the distances~$n \left( f / n \right)^\frac{1}{3}$ and~$n \left( f / n \right)^\frac{2}{3}$ along the projected view vector.
For each split, the polygon is clipped against a near plane and a far plane to define the required visible area inside a cascade.
For the first cascade, this near plane is~$n$, otherwise it is the far plane of the previous cascade.
The far plane for the last cascade is~$f$, otherwise it is the split distance.

To enable blending of values at cascade transitions, the far planes of the first and second cascades are offset by a small distance along the projected view vector so that polygons of adjacent cascades intersect.
Figure~\ref{fig:overview}b illustrates the resulting polygon (orange) for the last cascade of the frustum in Figure~\ref{fig:overview}a.
From the cascade polygons, axis-aligned bounding boxes are taken that have to be fitted into the square cascade textures.
Each bounding box should cover the largest possible area of the texture to utilize the texture resolution.
However, the bounding box defined in world space has to be offset as well to align the world space with the texture space.
If the offset of the world-space origin relative to the texture-space origin changes with the perspective, so do the texel locations -- and therefore the sample locations -- within adaptive grid cells.
This leads to temporal aliasing, which is most prominent in the last cascade.
For stable fitting, we widen the extents of each bounding box to be an even multiple of the adaptive grid's minimum cell size.
We then offset and scale the area of the bounding box within the texture such that all four corners of the bounding box align to texel centers of the texture.

The cascade textures are used to store and access the approximated heightfield data.
In the cascade texture illustrated in Figure~\ref{fig:overview}b, many texels are located outside the orange visible polygon.
Operations on these texels would be in vain and thus need to be avoided.
For this purpose, we maintain an additional texture per cascade to track whether a texel is inside the current cascade polygon.
In the following, we refer to texels inside the cascade polygons as visible texels.

\section{Surface Reconstruction with Discretized Radial Basis Functions}
\label{sec:sampling}

\begin{figure}[t]
	\centering
	\includegraphics[width=\linewidth]{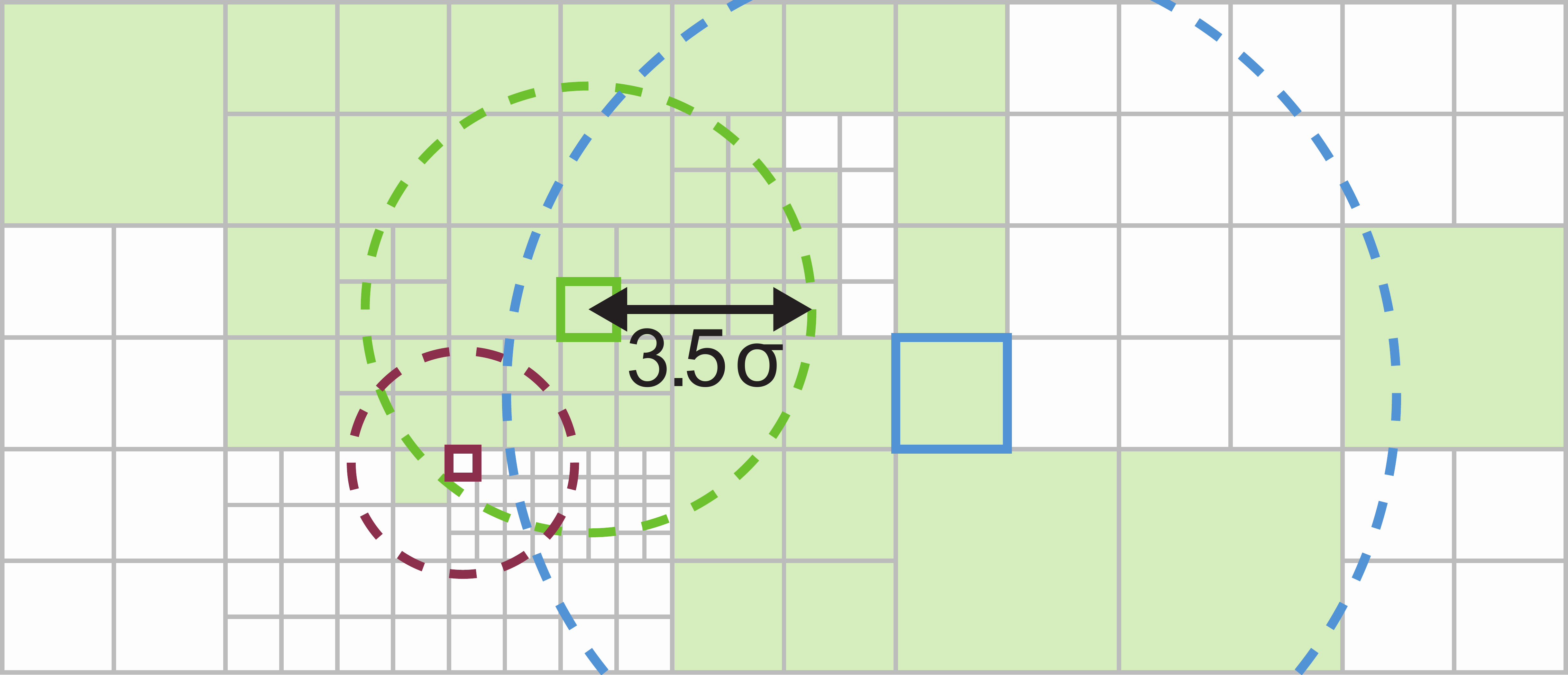}
	\caption{Influencing cells used for weighted averaging. The green cell influences any position inside the truncated radial basis function depicted by the dashed green circle. This analogously applies to the red and blue cells. All cells influencing the center of the green cell are filled with light green. Smaller cells covered by the green radial basis function might not influence this position~(red). Larger cells beyond the influence of the green radial basis function might still influence this position~(blue).}
	\label{fig:sampling_interp_gauss_cell_influences}
\end{figure}

To visualize discrete heightfield data as a continuous surface, a surface reconstruction from the given values at cell centers is needed.
While any higher-order interpolation or approximation is suitable for this task, we choose radial basis functions for their trivial parallelizability.
In particular, we use a Gaussian function as basis function for its desireable shape and controllable fall-off of influence.
The value of sample~$s$ at world-space position~$p_s$ is calculated from the set~$I$ of all influencing cells~$i$ of~$p_s$. A cell of size~$c_i$ with its center at~$p_i$ influences position~$p_s$ if the weight function~$w_i(p_s) > 0$.
This weight function is a Gaussian function truncated at $3.5 \sigma$, with the truncation remainder subtracted:
\begin{equation}
\label{eq:sampling_interp_gauss_weighted_sum_weight}
w_i(p_s) = \exp \left( -\frac{1}{2 \sigma^2} \left( \frac{\| p_i - p_s \|}{c_i} \right)^2 \right) - \exp \left( -\frac{3.5^2}{2} \right).
\end{equation}

The choice of~$\sigma$ influences quality and performance of the smoothing.
For larger values, more cells contribute to the value at~$p_s$, which results in a smoother approximation, but a higher computation time.
We chose~$\sigma = 1.0$, which is the smallest value that provides sufficiently smooth results over all considered real-world scenarios.

In Figure~\ref{fig:sampling_interp_gauss_cell_influences}, we illustrate the influences of three cells with three different sizes~$c_i$.
The influence of each cell is circular around the cell center and falls off exponentially until it is truncated at~$3.5\sigma$.
In this figure, the red cell lies within the dashed influence circle of the green cell, but its own influence is too small to contribute to the value at any position within the green cell.
Likewise, the green cell does not contribute to the blue cell, but values inside the green cell are still influenced by the blue cell.
All cells that influence the center position of the green cell are filled with a light green.
The set~$I$ of influencing cells at position~$p_s$ only depends on the adaptive grid, so it can be precomputed on startup.
Over the entire area of a cell, $I$ is likely to change very little, which is why we store all potentially influencing cells of each cell in a buffer for fast access.
Finally, the approximated heightfield value~$v_s$ of sample~$s$ at position~$p_s$ is the weighted sum of values~$v_i$ of all influencing cells~$i \in I$,
\begin{equation}
\label{eq:sampling_interp_gauss_weighted_sum}
v_s = \frac{\sum_{i \in I}{w_i(p_s) v_i}}{\sum_{i \in I}{w_i(p_s)}}.
\end{equation}
This sum is calculated for each texel of each of the three cascade textures in a compute shader.

\section{Rendering}
\label{sec:rendering}

After approximation, the calculated smooth data are available in the cached cascade textures in GPU memory that can be used for rendering.
Since our proposed approximation technique results in a regular heightfield, it is independent of the rendering technique used to display it.
We use ray casting with a maximum mipmap for acceleration as proposed by Tevs~et~al.~\cite{TevsEtAl2008}.
Our ray casting is implemented as a compute shader that casts one ray per pixel.
Each ray is traced trough all cascades, starting with the nearest one, and traversal stops upon ray-patch intersection or if the ray leaves the heightfield extents.
Within overlapping regions between two neighboring cascades, two hits will be evaluated.
The final intersection position is a convex combination of both intersections.

\section{Results}
\label{sec:results}

\begin{figure*}[ht]
	\centering
	\includegraphics[width=\linewidth]{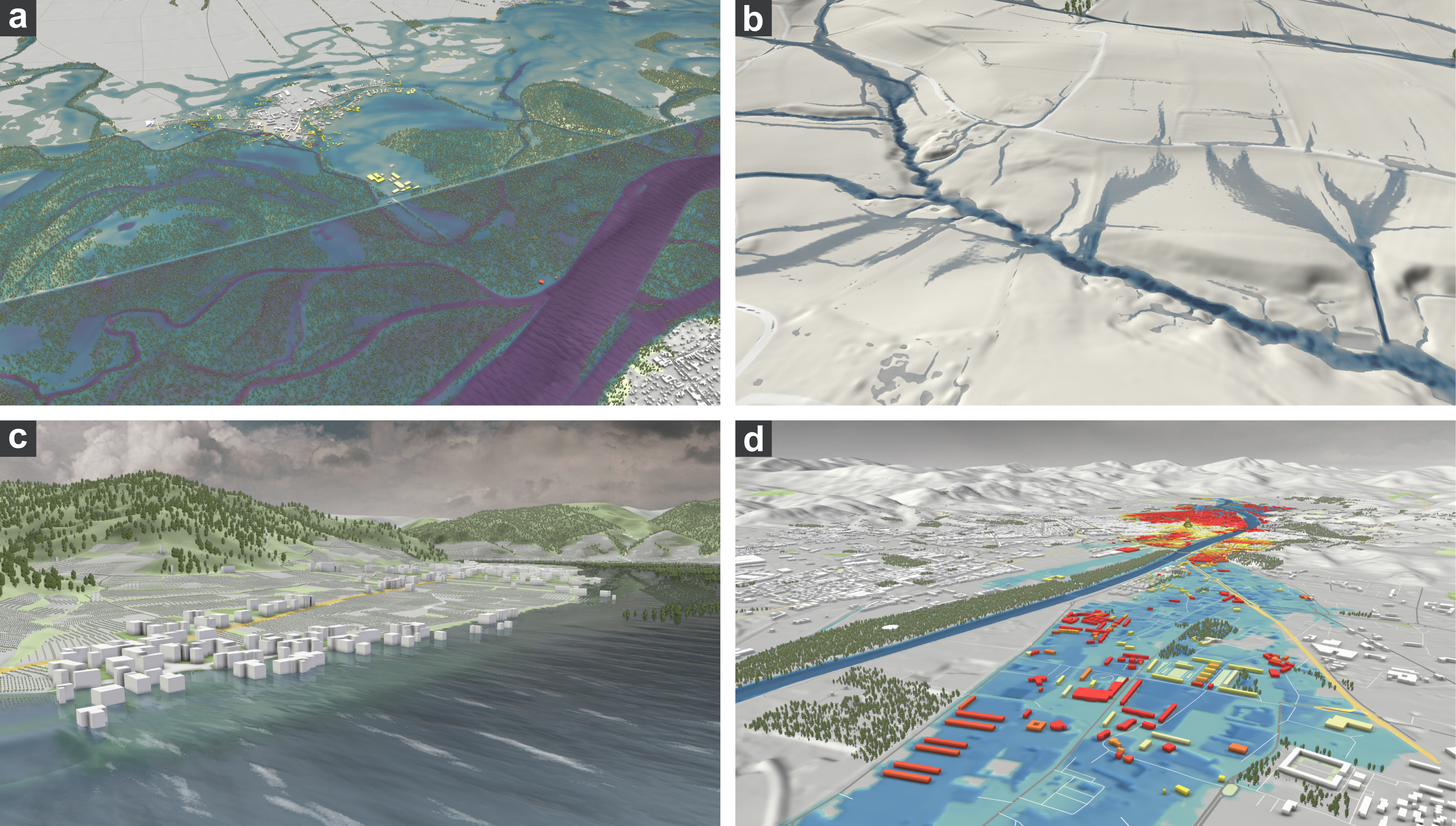}
	\caption{Real-world case studies. (a) Vast region in Marchfeld (Austria), a dyke break scenario. The rendering is efficient enough for interactive navigation. (b) Storm water scenario in Petzenkirchen (Austria) on a high-resolution domain. (c) A recent 100-year river flood in Wachau (Austria). (d) River flooding in Florence (Italy).}
	\label{fig:cases}
\end{figure*}

We evaluate the performance of our approach in four real-world scenarios.
These scenarios illustrate the different requirements with regards to scale, zoom range, and runtime performance that an interpolation method must meet in order to be used in a flood management software on a daily basis.
\newpage

\textbf{Marchfeld (Austria)} is a vast domain north of the Danube river, ranging from Vienna (Austria) to Bratislava (Slovakia).
A dyke stretching along the river for about 40~kilometers fences off the villages nearby.
Figure~\ref{fig:cases}a shows the five-day scenario of a dyke break during a once-in-a-century flood.
Interactive navigation enabled by our technique allows both to get an overview of the flooded region and to analyze the aftermath of the flood for individual villages or buildings.
Flood managers are thus able to create protection plans for the property and infrastructure.
In the figure, water depth is mapped with color onto the water surface.

\textbf{Petzenkirchen (Austria)} is an open-air hydrological laboratory.
A storm water runoff simulation of a very high resolution (Figure~\ref{fig:cases}b) delivers the results of 2~hours of a 48~mm/h precipitation.
The water surface color visualizes the magnitude of the local water velocities.

\textbf{Wachau (Austria)} is a valley on the Danube river and a part of Austria's cultural heritage.
A large-scale simulation shows the consequences of the once-in-a-century flood of May 2013 (Figure~\ref{fig:cases}c).

\textbf{Florence (Italy)}, situated on the Arno river, is our last case study.
River flooding in an urban setting has been simulated and the resulting inundation visualized in Figure~\ref{fig:cases}d.
Water depths are mapped onto the terrain, and the buildings are colored according to the water depths nearby.

Table~\ref{tab:resInput} contains the details of the presented scenarios.
Notice the mountainous part in the background, best visible in Figure~\ref{fig:cases}d.
This is a vast, low-resolution, decoration part of the domain that does not take part in the flood simulation but still needs to be rendered efficiently.

\begin{table} [ht]
  \begin{tabular*}{\linewidth}{|l @{\extracolsep{\fill}} |c|c|c|c|}
    \hline
&\enspace\textbf{Marchfeld}\enspace &\enspace\textbf{Petzenkirchen}\enspace &\enspace\textbf{Wachau}\enspace &\enspace\textbf{Florence}\enspace\\ \hline
\textbf{Extents [km]} & $82.9 \times 57.1$ & $15.4 \times 16.0$ & $24.1 \times 16.5$ & $44.6 \times 28.2$\\ \hline
\textbf{Cell Size [m]} & $3.00$ - $96$ & $0.75$ - $96$ & $3.00$ - $96$ & $3.00$ - $96$\\ \hline
\textbf{Cells in million} & $7.46$ & $1.81$ & $1.95$ & $4.53$\\ \hline
\textbf{Approximation [ms]\enspace\enspace} & $15.1$ & $20.8$ & $5.3$ & $1.5$\\ \hline
\textbf{Ray Casting [ms]} & $5.1$ & $4.3$ & $4.3$ & $2.8$\\ \hline
  \end{tabular*}
	\label{tab:resInput}
	\caption{Input information for the tested scenarios. The size of the visualized domain in kilometers, the range of the sizes for cells in meters, the amount of cells in millions, the duration of the heightfield approximation passes for terrain and water in milliseconds, and the duration of ray casting the heightfields in milliseconds.}
\end{table}

Our approach is implemented using OpenGL compute shaders and rendering.
Benchmarks were created on a system with an Intel I7-$3770$ $3.4$~GHz CPU, $32$ GB of RAM, and an Nvidia GTX~$1080$ GPU.
We used cascade textures of size $1024^2$ and a screen resolution of $1920\times1080$~pixels.
The last two columns of Table~\ref{tab:resInput} show the timings for the heightfield rendering in each scenario.

It can be seen that frame durations below 16~ms desired for real-time applications cannot always be achieved, based on the complexity of the scenario.
In particular, the approximation pass to discretize the radial basis functions needed for the efficient calculation of ray-surface intersections is very expensive.
Ray casting itself gives high quality results, but also occupies up to a third of the available time budget per frame, which we attribute to cache misses during ray traversal.
We want to emphasize the high frame duration in relation to the moderate number of cells in all of the scenarios.
In modern decision support systems, scenarios two magnitudes larger than the ones considered here are becoming the standard.
It is apparent from our benchmarks that the runtime performance of our approach does not scale well to 100 million cells and above.

\section{Conclusion}

With an increase in Earth observation activities and the increasing amount of raster data acquired, as well as with the significant advances in GPU-based environmental simulations, heightfield rendering in real time is becoming an integral part of more and more applications.
In this technical report, we discuss the direct rendering of heightfields defined on adaptive grids with higher-order interpolation using ray casting.
The discussed two-step approach allows us to decouple surface reconstruction from visualization, which then enables the use of complex interpolation or approximation that has previously been considered too slow for ray casting.
Still, we could not achieve satisfactory frame rates for moderately sized real-world scenarios due to the expensive necessary discretization of the smooth surface reconstruction and its caching.
We report our approach as a failed attempt to circumvent the limitations of ray casting that make it only suitable for simple, piece-wise polynomial surfaces on which a ray-surface intersection can be calculated efficiently. 
We conclude that for higher-order interpolation or approximation methods such as kriging, natural neighbor interpolation, local refinable splines, and thin-plate splines, triangulation-based heightfield rendering approaches are currently the only viable option for geospatial visualization in real time.

\section{Acknowledgments}

VRVis is funded by BMK, BMDW, Styria, SFG, Tyrol and Vienna Business Agency in the scope of COMET - Competence Centers for Excellent Technologies (879730) which is managed by FFG.

\end{document}